\documentclass[a4paper,12pt]{article}

\newcommand {\ket}[1]       { | #1 \rangle }

\newcommand {\braket}[2]    { \langle #1 | #2 \rangle }

\newcommand {\comment}[1]   {\Uparrow \mbox {\emph{#1}}}

\title {Exact solutions for a family of discretely spiked harmonic oscillators}
\author {Jan Skibi\'{n}ski \\ {\small Numeric Quest Inc., Huntsville, Ontario, Canada}\\
{\small jans@numeric-quest.com}}

\begin{document}
\maketitle
\tableofcontents
% ------------------------------
\section	{Introduction}
% ------------------------------
	
	The name \emph {spiked oscillator} generally refers to Hamiltonians
	with potentials containing a parabolic term $x^2$ describing
	ordinary quantum oscillator, and a singular term $1/x^\alpha$ forming
	a sharp spike at $x=0$.
	A typical Hamiltonian of a spiked oscillator is of the form given
	by Harrell\cite{Harrell}:
       \[ H = -\frac{d}{dx^2} + x^2 + \frac {\lambda}{x^\alpha} \qquad 0 \leq x < \infty. \]
	It has found variety of uses in atomic, molecular, nuclear and particle
	physics since it provides the simplest realistic model of interaction
	potentials due to its repulsive core $x^{-\alpha}$.

	Some forms of spiked oscillators have exact solutions, such as the one
	described by one dimensional Hamiltonian, examined four decades ago by Goldman
	and Krivchenkov\cite{Goldman}
	\[ H = -\frac{d}{d^2} + V_0(\frac{a}{x} - \frac{x}{a})^2 \]
  	whose exact solutions $\phi_n(x) \in [0,\infty)$
	satisfy Dirichlet's boundary condition $\phi_n(0)=0$.

	Models with exact solutions can serve as starting points
	for analyses of related cases where exact solutions do not
	exist and where a perturbative analysis is
	required\cite{Mustafa}. A good example can be found in the papers of
	Hall et al\cite{Hall}\cite{Hall2}, who slightly modified the above
	Hamiltonian, computed a basis given by a set of exact orthonormal
	eigenfunctions $\phi_n(x)$ for $L^2 (0,\infty)$, and then used
	this basis in variational analysis of the Hamiltonian
	\[ H = -\frac{d}{dx^2} + B x^2 + \frac{A}{x^2} + \frac{\lambda}{x^\alpha}. \]
	They argued that their singular basis forms much better starting
	point for perturbation techniques than the basis of the ordinary
	harmonic oscillator, as used in earlier work of Aguilera-Navarro
	et al\cite{Aguilera}, who had applied a perturbative variational
	analysis to the lowest eigenvalue of the Harrel's Hamiltonian.

 	In this context, we present an analysis of family of following
	spiked Hamiltonians

\begin{equation}
	H_l = -\frac{d^2}{dx^2} + x^2 + \frac{l(l+1)}{x^2}, \label{eq:hamiltonians}
\end{equation}
	parametrized by discrete values of $l = 0, 1, 2..$.
        Depending on interpretation of variable $x$, this form can either
	describe a family of one dimensional linear oscillators, or a family of
	$N$ dimensional radial oscillators for odd $N=1,3,5..$
	What is traditionally known as the three dimensional isotropic radial
	oscillator, with known exact solutions, can be reduced to a special
	case of $N=1$. We shall show that solutions for
	radial oscillators in even dimensions $N=2,4,6..$ do not form
	well defined orthogonal bases and, as such, they are of no practical interest.

	The solutions we provide are exact and much simpler
	than any of those described in this introduction, and therefore they
	are good candidates for orthonormal bases that could be used
	in perturbative analyses of oscillators with spiked potentials other
	than $x^{-2}$.

	The factorization formalism developed here is completely algebraic
	and somehow related to methods of \emph{Superstring Quantum Mechanics (SUSY QM)}.
 	Modern algebraic factorization methods are overviewed in
	the paper of Rosas-Ortiz\cite{Ortiz} and closely related SUSY methods
	are described for example in introductory parts of the papers by Junker and Roy
	\cite{Junker}\cite{Junker2}, and L\'evai and Roy\cite{Levai}.

	The common premise of all those methods is the use of algebraic
	recursive manipulations on two Hamiltonians: one with known
	spectrum and known eigenvectors and another which is isospectral
	to the former but whose eigenvectors are to be found. The methodology
	developed here uses the same premise, but goes a bit further by
	manipulating not just two Hamiltonians but the uncountable set of related
	Hamiltonians (\ref{eq:hamiltonians}). On the other hand, the SUSY and similar methods rely
	on general solution of underlying \emph{Ricatti} equation, which is
	completely ignored in the approach presented in this paper.

	Quantum models with exact solutions play an important role and there
	is a renewed interest in enrichment of the traditional narrow set
	of exactly solvable models, such as harmonic oscillator, hydrogen-like potential, Morse
	potential and square well potential. It is only fair to mention other
	related methods, which admit certain type
	of non-hermitian Hamiltonians with complex-valued potentials giving
	rise to real energy spectrum. For example,
	Cannata et al\cite{Cannata} use Darboux method to derive a general
	class of complex potentials with energy spectra identical
	to that of regular harmonic oscillator.
	Bender et al\cite{Bender}\cite{Bender2} replaces the requirement of
	Hamiltonian hermiticity by weaker requirement of \emph{PT-symmetry},
	which is satisfactory condition for real energy spectra. They
	examine complex harmonic and anharmonic oscillators, complex
	square wells, etc. This avenue is being explored by many other
	researchers.
	
	To complete this introduction, we just mention in passing
	that exactly solvable problems can be categorized in one
	of three classes: \emph{Exactly solvable (ES), Quasi exactly
	solvable (QES) and Conditionally exactly solvable (CES)}.

	In the next section we present an overview of
	our findings, while the remaining sections provide specific details.

% --------------------------------
\section     {Outline}
% -------------------------------
	The first Hamiltonian, $H_0$, from the family of discretely spiked
	oscillators (\ref{eq:hamiltonians}) describes the well known harmonic oscillator,
	without the spike, of known discrete energy spectrum and known ladder
	operators $a_0$ and $a_0^\dag$.
	Aside from this simple case with parabolic potential, the potentials of remaining
	members of this family have strong discontinuities at $x = 0$ and
	do not resemble the first case at all. Yet
	they all are isospectral; that is, they all have the discrete energy
	spectra identical to that of the simple harmonic oscillator. The only
	difference is in the definition of their ground state energies,
	$E_{l,0}=2l+1$, in units of $\frac{1}{2}\omega\hbar$.

	We have chosen to denotate the coresponding  ground states as
	$\ket {0,0}$, $\ket {1,0}$, $\ket {2,0}$, etc. for $l=0,1,2..$, respectively.
	Table \ref{states} visualizes the relationship between energy states
	and energies for the entire family of discretely spiked oscillators.

\begin{table}

\caption{States and energies of discretely spiked harmonic oscillators. Some states
are unphysical and will be later rejected.}

\label{states}
\vspace {2 mm}
\begin{quote}
\begin{quote}
\begin{tabular}{|l|l|l|l|l|}
\hline
	$l = 0$         &$l = 1$           &$l = 2$          &$l =$...   &Energy       \\
\hline
        ...             &...               &...              &...        &...          \\
	...             &...               &$\ket {2,k}$     &...        &$2k+5$     \\
	...             &$\ket {1,k}$      &$\ket {2,k-1}$   &...        &$2k+3$     \\
	$\ket {0,k}$    &$\ket {1,k-1}$    &$\ket {2,k-2}$   &...        &$2k+1$     \\
	$\ket {0,k-1}$  &$\ket {1,k-2}$    &...              &...        &$2k-1$     \\
        $\ket {0,k-2}$  &...               &...              &...        &$2k-3$     \\
	...             &...               &...              &...        &...          \\
        ...             &...               &$\ket {2,2}$     &...        &9            \\
	...             &$\ket {1,2}$      &$\ket {2,1}$     &...        &7            \\
        $\ket {0,2}$    &$\ket {1,1}$      &$\ket {2,0}$     &\bf-----   &5            \\
        $\ket {0,1}$    &$\ket {1,0}$      &\bf-----         &\bf-----   &3            \\
	$\ket {0,0}$    &\bf-----          &\bf-----         &\bf-----   &1            \\
\hline
\end{tabular}
\end{quote}
\end{quote}
\end{table}

	For each Hamiltonian $H_l$ a pair of intertwining operators can
	be defined
\begin{eqnarray*}
 	b_{l+1}^\dag \ket {l,k}     & \propto & \ket {l+1,k} \\
	b_{l+1}      \ket {l+1,k} & \propto & \ket {l,k}
\end{eqnarray*}
        which connect eigenstates of two `neighbours', $H_l$ and
	$H_{l+1}$. Both represent actions coresponding to diagonal traversal
	of the Table \ref{states}, in NE and SW directions, respectively.

	In addition to intertwining operators, we can also define
	the ladder operators $a_l$ and $a_l^\dag$ that allow for traversing
	of the Table \ref{states} in vertical direction.
	The annihilation operator changes the state $\ket {l,k}$ with energy
	$E_{l,k}$ to the state $\ket {l,k-1}$ of lower energy $E_{l,k-1}$
\begin{eqnarray*}
	a_l      \ket {l,0} & = & 0 \\
	a_l      \ket {l,k} & = & \alpha_l \ket {l,k-1}.
\end{eqnarray*}
	Similarly, the creation operator changes the state $\ket {l,k}$ with
	energy $E_{l,k}$ to the state $\ket {l,k+1}$ of higher energy $E_{l,k+1}$
	\[ a_l^\dag \ket {l,k} = \alpha_l'  \ket {l,k+1} \]
	where $\alpha_l$ and $\alpha_l'$ are the normalization constants.

	Known ladder operators $a_0$ and $a_0^\dag$ for the classical
	harmonic oscillator are used to recursively prove the existence
	of similar operators for the remaining Hamiltonians $H_l$.
	The are given by formulas
	\[ a_{l+1}^\dag = b_{l+1}^\dag a_l^\dag b_{l+1} \]
	\[ a_{l+1}      = b_{l+1}^\dag a_l b_{l+1}      \]
	
	The condition
	$a_l \ket {l,0} = 0$ allows to generate explicit form of the state $\ket {l,0}$
	in position representation; that is the vaweform function $\phi_{l,0}(x)$ of the
	ground state.
	For $l = 0$, this is the bell-shaped function, $\phi_{0,0} \propto \exp {(\frac{-x^2}{2})}$, subject
	to normalization. Having computed the ground waveform $\phi_{l,0}$, one can use
	the creation operator $a_l^\dag$ to recusively generate all other eigenstates
	$\ket {l,k}$, or coresponding waveforms $\phi_{l,k}(x)$.

	This seemingly completes the scheme since, thanks to the
	pairs of ladder and intertwining operators, the entire
	Table \ref{states} can be created -- starting from the left lower corner and zigzagging
	one step a time, with or without the help of ladder operators $a_1^\dag$,
	$a_2^\dag$, etc.

	However, we must also require that each set of state
	vectors is a basis in its own space; that is, they all are mutually orthogonal
	and normalized
	\[ \braket {l,j} {l,k} = \delta_{jk} .\]
	But some of the wavefunctions are singular at $x=0$, since they
	contain factors $\frac{1}{x}$, $\frac{1}{x^2}$, etc. Depending on
	whether the Hamiltonians (\ref{eq:hamiltonians})
	describe linear or radial oscillators and, in the latter case,
	depending on the space dimension, some such functions are not square-integrable
	and therefore they must be rejected as unphysical. Those wave functions that
	are physical must additionally pass the test of orthogonality in order
	to qualify as members of useful bases.
	To take it all into account we therefore specialize the generic scheme and
	draw additional conclusions specific to dimensionality of the problem.
	All of this is supported by a Haskell program, which tests the theory
	and provides useful and flexible computational tool.
	
	In the following sections we shall present a detailed account of this
	outline.	

% -----------------------------------
\section {Operators}
% -----------------------------------

	We shall define two pairs of state generation operators: intertwining
	operators, affecting the quantum number $l$, and ladder operators affecting
	the qunatum number $k$. The former correspond to diagonal traversal of
	Table \ref{states} and the latter -- for the vertical traversal.

% -----------------------------------
\subsection {Intertwining operators}
% -----------------------------------

	Let's define two operators $b_l$ and its adjoint $b_l^\dag$
\begin{eqnarray}
	b_l      &=&  \frac{d}{dx} + \beta_l(x), \\
	b_l^\dag &=& -\frac{d}{dx} + \beta_l(x) \qquad \textrm{where} \\
	\beta_l(x) &=& x + \frac {l}{x}.
\end{eqnarray}

	Function $\beta(x)$ and operator $\frac{d}{dx}$ do not commute

	\[ [\beta_l, \frac{d}{dx}] = -\beta_l'(x)  = \frac{l}{x^2} - 1,\]
	and, as a consequence, the commutator $[b_l,b_l^\dag]$ is not zero
	either, since
\begin{eqnarray}
	b_l^\dag b_l &=& -\frac{d^2}{dx^2} + \beta_l^2(x) - \beta_l'(x)
                        = H_l + (2l-1), \label{bdagb}\\
	b_l b_l^\dag &=& -\frac{d^2}{dx^2} + \beta_l^2(x) + \beta_l'(x)
                        = H_{l-1} + (2l+1),
	\label{eq:bbdag}
\end{eqnarray}
	where
	\[ H_{-1} = H_0 = -\frac{d^2}{dx^2} + x^2 ,\]
	and where the remaining Hamiltonians are defined by (\ref{eq:hamiltonians}).

	With the help of the above equations we can establish two very
	useful formulas, which we will use later:
\begin{eqnarray}
	H_l b_l^\dag - b_l^\dag H_{l-1} &=& 2b_l^\dag, \label{eq:hbdag} \\
	H_{l-1} b_l - b_l H_l &=& -2b_l.               \label{eq:hb}
\end{eqnarray}

	These are quasi commutative relations, connecting two ``neighbouring''
	Hamiltonians. Let us now consider specific cases of $\beta_l(x)$.

% --------------------------------
\subsubsection {Ordinary oscillator}
% -------------------------------

	The case $l=0$ is very special since
\begin{eqnarray*}
	b_0^\dag b_0 &=& H_0 - 1, \\
	b_0 b_0^\dag &=& H_{-1} + 1\\
                     &=& H_0 + 1 .
\end{eqnarray*}

	Both equations refer to the same Hamiltonian of ordinary harmonic
	oscillator, and the operators $b_0$ and $b_0^\dag$ degenerate to
	the familiar ladder operators
\begin{eqnarray}
	a_0^\dag &=& b_0^\dag = -\frac{d}{dx} + x, \\
	a_0      &=& b_0      =  \frac{d}{dx} + x,
\end{eqnarray}
	which have these important properties:

\begin{eqnarray}
        \lbrack a_0, a_0^\dag] & = & 2,       \\
        \lbrack H_0, a_0]      & = & -2a_0,   \\
        \lbrack H_0, a_0^\dag] & = & 2a_0^\dag.
\end{eqnarray}

	Consequently, the energies are given by $E_{0,k} = 2k+1$ for $k=0,1,2$...,
	in units of $\frac{1}{2}\omega\hbar$. We will skip other details,
	since this is the very known case.

% -----------------------------------------
\subsubsection {Discretely spiked oscillators}
% -----------------------------------------

	The interesting part begins here. First, let us prove that
	$b_{l+1}^\dag \ket {l,k}$ is an eigenvector of Hamiltonian $H_{l+1}$
	by assuming that this eigen equation holds
	\[ H_l \ket {l,k} = E_{l,k} \ket {l,k} \]
	and making use of formula (\ref{eq:hbdag}):
\begin{eqnarray*}
	H_{l+1} \ket {b_{l+1}^\dag (l,k)} & = & H_{l+1} b_{l+1}^\dag \ket{l,k} \\
           & = & b_{l+1}^\dag H_l \ket {l,k} + 2b_{l+1}^\dag \ket{l,k} \\
	   & = & (E_{l,k} + 2) \ket {b_{l+1}^\dag (l,k)}.
\end{eqnarray*}
	The operator $b_{l+1}^\dag$ transforms the state $\ket{l,k}$ with energy
	$E_{l,k}$ into the state $\ket {l+1,k}$ with energy $E_{l+1,k} = E_{l,k} + 2$

\begin{equation}
	\ket {l+1,k} = \frac {1} {\sqrt {2k+4l+4}} b_{l+1}^\dag \ket {l,k}.
	\label{eq:moveNE}
\end{equation}

	This coresponds to the diagonal move in NE direction in Table \ref{states}.
	One can verify the normalization factor in equation (\ref{eq:moveNE}) by
	comparing the norm of its both sides, making use of (\ref{eq:bbdag}) and noticing
	that the energy of l-th oscillator in the state $\ket {l,k}$
	is given by
\begin{equation}
	E_{l,k} = 2 (l + k) + 1.   \label{eq:energies}
\end{equation}
	The latter can be easily established with the help of Table \ref{states}.

	Similarly, we can prove that state $b_{l+1} \ket {l+1,k}$ is
	the eigenvector of Hamiltonian $H_l$:
\begin{eqnarray*}
	H_l (b_{l+1} \ket {l+1,k}) & = & (b_{l+1} H_{l+1} - 2 b_{l+1}) \ket {l+1,k})\\
	                             & = & (E_{l+1,k} - 2) (b_{l+1} \ket {l+1,k}).
\end{eqnarray*}

	The operator $b_{l+1}$ transforms the state $\ket{l+1,k}$ with energy
	$E_{l+1,k}$ into the state $\ket {l,k}$ with energy $E_{l,k} = E_{l+1,k} - 2$
\begin{equation}
	\ket {l,k} = \frac {1} {\sqrt {2k+4l+4}} b_{l+1}\ket {l+1,k}.
\end{equation}
	This coresponds to the diagonal move in SW direction in Table \ref{states}.

% ----------------------------------
\subsection {Ladder operators}
% -----------------------------------

	We shall now define two operators
\begin{equation}
	a_{l+1}^\dag = b_{l+1}^\dag a_l^\dag b_{l+1} \label{eq:aladderdag}
\end{equation}
	and
\begin{equation}
	a_{l+1}      = b_{l+1}^\dag a_l b_{l+1} \label{eq:aladder}
\end{equation}
	and prove that they indeed have the expected properties of ladder
	operators. We shall first prove by induction, that
\begin{equation}
	[H_l, a_l^\dag] = 2a_l^\dag  \label{eq:haladderdag}
\end{equation}
	This holds true for the case $l=0$, as seen in (13). Assuming now,
	that the above is true for $l-1$, the proof is as follows:
\begin{eqnarray*}
	H_l a_l^\dag & = & H_l b_l^\dag a_{l-1}^\dag b_l \\
	         & & \comment{from \ref{eq:aladderdag}} \\
	             & = & (b_l^\dag H_{l-1} + 2b_l^\dag) a_{l-1}^\dag b_l \\
		 & & \comment{from \ref{eq:hbdag}} \\
	             & = & b_l^\dag (a_{l-1}^\dag H_{l-1} + 2a_{l-1}^\dag) b_l + 2a_l^\dag \\
		 & & \comment {from \ref{eq:haladderdag} by induction} \\
	             & = & b_l^\dag a_{l-1}^\dag H_{l-1} b_l + 4a_l^\dag  \\
	             & = & b_l^\dag a_{l-1}^\dag (b_l H_l - 2b_l) + 4a_l^\dag  \\
		 & & \comment{from \ref{eq:hb}} \\
	             & = & b_l^\dag a_{l-1}^\dag b_l H_l + 2a_l^\dag \\
	             & = & a_l^\dag H_l + 2a_l^\dag  \\
		 & & \comment{from \ref{eq:aladderdag}}
\end{eqnarray*}

	Similarly, we can prove by induction that the following holds
\begin{equation}
	[H_l, a_l] = -2a_l.  \label{eq:haladder}
\end{equation}
	It immediately follows from (\ref{eq:haladderdag}) and (\ref{eq:haladder}) that operators $a_l^\dag$
        and $a_l$ create and annihilate eigenstates of Hamiltonian $H_l$, respectively
\begin{eqnarray*}
	H_l a_l^\dag \ket {l,k} & = & a_l^\dag H_l \ket {l,k} + 2 a_l^\dag \ket {l,k}\\
                                & = & (E_{l,k} + 2) a_l^\dag \ket {l,k},
\end{eqnarray*}

\begin{eqnarray*}
	H_l a_l \ket {l,k} & = & a_l H_l \ket {l,k} - 2 a_l \ket {l,k} \\
                           & = & (E_{l,k} - 2) a_l \ket {l,k}.
\end{eqnarray*}
	Finally, we can prove by induction that if $a_0 \ket {0,0} = 0$
	then similar relation holds for any other oscillator ($l > 0$):
	\[ a_l \ket {l,0} = 0 .\]

% ------------------------------------------------
\section {Wavefunctions}
% -----------------------------------------------

% -----------------------------------------------
\subsection {Haskell module Spike}

	The remaining part of this paper refers to results obtained
	with the help of a little program,\emph{Spike.hs},\footnote
	{The program is available at http://www.numeric-quest.com/haskell/Spike.hs.}
	written in functional lazy language Haskell\footnote{See home page of
	Haskell at http://www.haskell.org.}. The program
	is used for generation of wavefunctions, for normalization of solutions
	in n-dimensional spaces and,
	generally, for testing the theory presented in the previous sections.
	The program performs some algebraic manipulation of Laurent series,
	such as computation of derivatives, integrals, sums
	and products, and directly implements raising and intertwining
	operators previously discussed. It is best used in interpreting
	environment\footnote{Haskell interpreter Hugs is available for free
	at http://www.haskell.org/hugs/. It runs on any major platform.}, since one can evaluate any formula
	from module \emph{Spike} in any order of one's choice.
	Module \emph{Spike} relies on other standard Haskell modules and on our
	module \emph{Fraction.hs} \footnote{Module Fraction is avauilable at
	http://www.numeric-quest.com/haskell/Fraction.hs.}.
	For interesting introduction to Haskell as a coding tool for scientific
	applications see\cite{Karczmarczuk}.

% -------------------------------------------------
\subsection {Laurent series}
% -------------------------------------------------

	Any analytic function $f(z)$ in an open annulus $r < |z-z_0| < R$
	can be expressed as the sum of two series
	\[ f(z) = \sum_{j=0}^{\infty} a_j(z-z_0)^j + \sum_{j=1}^{\infty} a_{-j} (z-z_0)^{-j}.\]
	Such an expansion, containing negative as well as positive powers is called
	\emph{Laurent series} for $f(z)$ in this annulus. In Haskell we will
	represent it as a pair of two lists
	\[ f = L \, [a_0,a_1,a_2,...] \, [a_{-1},a_{-2},...], \]
	where coefficients of expansions can be any numbers belonging to class
	\emph{Num}, such as \emph{Integer, Double, Complex Double}, etc.

The following
	defines the type of data structure representing Laurent series
\begin{verbatim}
data Laurent a = L [a] [a],
\end{verbatim}
	where $a$ is a type variable of some numeric kind. It is convenient
	to choose it as $Integer$ (of unlimited size) since this leads to highly
	accurate results. For computations requiring normalization the data
	\emph{Fraction} will be used instead of traditional \emph{Double}
	since this guarantees very good accuracy, with relatively small
	performance loss.

	As it will soon become clear, wavefunctions of spiked oscillators can be
	represented as products of two functions
\begin{equation}
	\phi_{l,k}(x) = f_{l,k}(x) e^{-x^2/2} \label{eq:solution}
\end{equation}
	where $f_{l,k}(x)$ is a real function of $x$ and can be expanded
	as Laurent series of type $Laurent \, Integer$.

% -----------------------------------------------
\subsection {One step operators}
% -----------------------------------------------

	Application of the intertwining operator to function (\ref{eq:solution})
\begin{eqnarray*}
	b_l^\dag \phi_{l,k}(x) & = & b_l^\dag (f_{k,l}(x) e^{-x^2/2}) \\
                         & = & (-f_{k,l}'(x) + 2x f_{k,l}(x) + \frac{l}{x} f_{k,l}(x)) e^{-x^2/2}\\
                         & = & (-f_{k,l}'(x) + p_l(x) f_{k,l}(x)) e^{-x^2/2},
\end{eqnarray*}
	results in manipulation of two Laurent series $p_l$ and $f_{k,l}$,
	where $p_l(x)$ is a short Laurent series, represented in Haskell
	as $L \, [0,2] \, [l]$. The expression $(-f_{k,l}'(x) + p_l(x) f_{k,l}(x)$
	is implemented in Haskell as function \emph{b'}. It takes as an input an integer
	$l$ and a Laurent series $f_{l,k}$ of type $Laurent \,Integer$,
	and then produces another Laurent series by invoking few primitives,
	such as multiplication, addition and differentiation defined
	for such series. Notice the \emph{gcd normalization} of the results,
	which keeps the size of integers in check and also provides
	a canonical standard for comparison of different methods of computations.
\begin{verbatim}
b'     :: Integer -> Laurent Integer -> Laurent Integer
b' l f = gcdNormalize (p * f - diff f)
    where
        p = L [0,2] [l]	
\end{verbatim}
	Implementation $b' \, l$ of operator $b_l^\dag$ is the only Haskell
	function needed to generate all eigenfunctions for all spiked
	harmonic oscillators. But to complement the theory developed
	in the previous sections, below are the remaining operators.
	Firstly, here is function $b \,l$ coresponding to intertwining operator $b_l$
\begin{verbatim}
b     :: Integer -> Laurent Integer -> Laurent Integer
b l f = gcdNormalize (p * f + diff f)
    where
        p = L [] [l].
\end{verbatim}
	Next is the implementation of the rasing operator $a^\dag$:
\begin{verbatim}
a' :: Integer -> Laurent Integer -> Laurent Integer
a' 0 = b' 0
a' l = b' l . a' (l - 1) . b l
\end{verbatim}
	Notice that function $a' \, 0 = b' \, 0$ is a special case. It coresponds to
	the raising operator $a_0^\dag = b_0^\dag$ -- exactly as specified
	in the theoretical sections. Finally, here is the implementation of the
	lowering ladder operator $a_l$:	
\begin{verbatim}
a :: Integer -> Laurent Integer -> Laurent Integer
a 0 = b 0
a l = b' l . a (l - 1) . b l.
\end{verbatim}

% ----------------------------------------------------
\subsection {Cumulative operators}

	Since the one-step operators from the previous section are
	to be applied recursively to some base functions it is convenient
	to define cumulative versions of these operators.

	The cumulative intertwining operator,\emph{twine' l}, is a composition
	of (NE) operators $b_l\dag \, b_{l-1}\dag \,...\,b_1\dag$

\begin{verbatim}
twine'   :: Integer -> Laurent Integer -> Laurent Integer
twine' 0 = id
twine' 1 = b' 1
twine' l = b' l . twine' (l - 1)
\end{verbatim}

	The cumulative raising ladder operator, \emph{ladder' l k}, is a composition
	of $k$ raising ladder operators $a_l^\dag \, a_l^\dag \, a_l^\dag ...$

\begin{verbatim}
ladder' :: Integer -> Integer -> Laurent Integer
           -> Laurent Integer
ladder' l 0 = id
ladder' l 1 = a' l
ladder' l k = a' l . ladder' l (k - 1).
\end{verbatim}

% -----------------------------------------------
\subsection {Generating wavefunctions}
% -----------------------------------------------

	Using these tools we can easily generate sets of eigenfunctions
	for spiked oscillators. When $l=0$ the first wavefunction
	is given by $e^{-x^/2}$ and that implies that $f_{0,0}$
	is simply $1$, and its corresponding Laurent series is
	$L\,\lbrack 1\rbrack\,\lbrack\rbrack$. Here  $\lbrack\rbrack$ represents
	an empty list -- meaning no negative powers at all.
	By executing
\begin{verbatim}
iterate (a' 0) (L [1] [])
\end{verbatim}
	we can generate infinite list of gcd-normalized eigenfunctions
	for ordinary harmonic oscillator. If we want to retain six, say,
	wave functions we need to cut this infinite list down to six elements

\begin{verbatim}
take 6 (iterate (a' 0) (L [1] [])).
\end{verbatim}

A slightly prettiefied output of the above looks like this:
\begin{verbatim}
[                           -- Meaning:
  L [1]                 [], -- 1
  L [0,1]               [], -- x
  L [-1,0,2]            [], -- -1  + 2x^2
  L [0,-3,0,2]          [], -- -3x + 2x^3
  L [3,0,-12,0,4]       [], --  3  - 12x^2 + 4x^4
  L [0,15,0,-20,0,4]    []  -- 15x - 20x^3 + 4x^5.
]
\end{verbatim}
	It is evident that these solutions do not contain negative powers
	of $x$. In fact they represent (gcd normalized) Hermite polynomials in disguise,
	and this is exactly what one would expect from ordinary harmonic
	oscillator. To test implementation of the lowering ladder operator
	$a_0$, execute
\begin{verbatim}
take 6 (iterate (a 0) (L [0,15,0,-20,0,4])).
\end{verbatim}
	This should recreate the above list in reverse. In addition,
	the expression
\begin{verbatim}
a 0 (L [1] [])
\end{verbatim}
	generates a special kind of Laurent series
\begin{verbatim}
L [] [],
\end{verbatim}
	which is the Laurent series for \emph{function zero}.

	To find eigenfunctions of the first spiked oscillator ($l=1$)
	we have two choices. Firstly, we can apply operator $b_1^\dag$, or
	function $b' \, 1$ to every eigenfunction
	of ordinary oscillator, by executing
\begin{verbatim}
map (b' 1) (take 6 $ iterate (a' 0) (L [1] [])).
\end{verbatim}
	Alternatively, we can iteratively apply function $a' \, 1$
	to the first eigenfunction $f_{1,0} = b \, 1 \, (L \, [1] \,[])$
	of the oscillator $l=1$
\begin{verbatim}
take 6 $ iterate (a' 1) f where f = b' 1 (L [1] []).
\end{verbatim}
	Either way, the output is as follows	
\begin{verbatim}
[                               -- Meaning:
  L [0,2]                 [1],  -- 2x + 1/x
  L [0,0,1]               [],   --  x^2
  L [0,-4,0,4]            [-1], -- -4x   + 4x^3   - 1/x
  L [0,0,-5,0,2]          [],   -- -5x^2 + 2x^4
  L [0,18,0,-36,0,8]      [3],  -- 18x   - 36x^3  + 8x^5 + 3/x
  L [0,0,35,0,-28,0,4]    []    -- 35x^2 - 28x^4  + 4x^6
]
\end{verbatim}
	This time, every other solution contains term with $1/x$, and as
	such it might or might not be acceptable as physical solution,
	depending on the space dimension under consideration. More
	about it in later sections.

	To generate similar list for the second spiked oscillator ($l=2$)
	we need to apply two functions $b' \, 1$ and $b' \, 2$ in
	succession to a list of eigenfunctions of ordinary harmonic
	oscillator
\begin{verbatim}
map (b' 2 . b' 1) (take 6 $ iterate (a' 0) (L [1] [])),
\end{verbatim}
	or, alternatively, apply the raising ladder function $a' \, 2$
	to the lowest eigenfunction of the oscillator $l=2$
\begin{verbatim}
take 6 $ iterate (a' 2) f where f = (b' 2 . b' 1) (L [1] []).
\end{verbatim}
	Both methods lead to the same output:
\begin{verbatim}
[
  L [4,0,4]                    [0,3],
  L [0,0,0,1]                  [],
  L [-6,0,-12,0,8]             [0,-3],
  L [0,0,0,-7,0,2]             [],
  L [24,0,72,0,-96,0,16]       [0,9],
  L [0,0,0,63,0,-36,0,4]       []
].
\end{verbatim}	
	Here is a general formula for generation of infinite lists of wavefunctions
	for spiked harmonic oscillators $l = 0,1,2..$
\begin{verbatim}
wavefunctions   :: Integer -> Laurent Integer
wavefunctions l = map (twine' l) $ iterate (a' 0) (L [1] [])
\end{verbatim}
	To generate just one eigenfunction $f_{l,k}$ the following Haskell
	function can be used
\begin{verbatim}
wavefunction     :: Integer -> Integer -> Laurent Integer
wavefunction l k = twine' l (ladder' 0 k (L [1] []))
\end{verbatim}

% ---------------------------
\section {Normalization and orthogonalization}
% ---------------------------
        We have shown that the solutions for spiked
	oscillators are of the form $\phi_{l,k}=f_{l,k}(x)e^{-x^2/2}$,
	where $f_{l,k}(x)$ are real functions of $x$ and, specificly, they are some
	Laurent series -- admitting both positive and negative
	powers of $x$. For a specific case of ordinary oscillator, these
	are the Hermite polynomials, with positive powers only.

	The scheme presented so far accepts all solutions, but does not check
	whether they are square-integrable and, in addition, whether they obey
	the orthogonality relations
	\[ \braket {l,i}{l,j} = \delta_{i,j}. \]
        Due to presence of the term $\frac{1}{x}$ in the definition of $b_l^\dag$
	some of the wavefunctions $\phi_{l,k}$ are tainted by powers of
	$\frac{1}{x}$ and as such might not be admissible as physical solutions
	for a dimension under consideration. Terms such as
	\[ \frac{1}{x} e^{-x^2/2}, \]
	when considered in one dimension, imply normalization integrals such as
	\[ \int_{-\infty}^{+\infty} \frac{1}{x^2} e^{-x^2} \, dx\]
	which are not integrable. On the other hand, if variable $x$ is
	treated as three dimensional radius then the normalization integral
	like this
	\[ \int_0^{+\infty} \frac{1}{x^2} e^{-x^2} 4\pi x^2 \,dx, \]
	is integrable.

% --------------------------------
\subsection {Admissible physical solutions}
% --------------------------------

	A general pattern of physically acceptable solutions is as follows:
\begin{itemize}
\item All odd eigenfunctions $\phi_{l,k} \, \mbox{for } k=1,3,5...$ of any oscillator
	$l=0,1,2...$ are physically acceptable, regardless the space dimension.
\item All even eigenfunctions $\phi_{l,k} \, \mbox{for } k=0,2,4...$ of any oscillator
	$l=0,1,2...$ are acceptable as physical solutions only when
	$N \geq 2l + 1$, where $N$ is the space dimension.
\end{itemize}
	The above statements can be verified by  	
	Haskell function \emph{physicalPattern}, which produces infinite list
	of booleans -- each stating whether a solution $k$ for oscillator $l$
	and space dimension $N$ is admissible as physical solution
\begin{verbatim}
physicalPattern :: Integer -> Integer -> [Bool]
physicalPattern = map (isPhysical n) (wavefunctions l)

isPhysical :: Integer -> Laurent Integer -> Bool
isPhysical n f
    | length us' == 0 = True
    | otherwise       = False
    where
        L us us' = f * f * dV
        dV = diff $ L ((take (fromIntegral n)
             $ repeat 0)++[1]) []
\end{verbatim}
	
	For example, even solutions for oscillator $l=2$ are unphysical
	in three dimensional space, $N=3$, as seen below (Haskell lists
	are traditionally indexed from zero)
\begin{verbatim}
take 6 $ physicalPattern 3 2
==> [False,True,False,True,False,True].
\end{verbatim}
	
% -------------------------------------------------
\subsection	{Linear normalization}
% -------------------------------------------------

	In order to normalize the eigenfunctions of one dimensional
	linear spiked oscillators and to test their orthogonality conditions we need
	algebraic formulae for the values of improper integrals
	
	\[P_n = \int_{-\infty}^{+\infty} x^n \, e^{-x^2} \, dx. \]
	They are as follows

\begin{eqnarray*}
        P_0 & = & \sqrt{\pi} \\
        P_1 & = & 0 \\
        P_n & = & \frac{n-1}{2} \, P_{n-2}.
\end{eqnarray*}
	Notice that integrals with odd powers of $x^n$ are all zero.

	All solutions for the ordinary harmonic oscillator ($l=0$) are
	physically acceptable since their Laurent series do not
	contain coefficients of negative powers of $x$. What's more, they form
	the orthogonal system of eigenfunctions

	\[ \braket {0,k} {0,k'} = \delta_{kk'} \]

	Physically  acceptable solutions for the remaining, spiked,
	oscillators ($l > 0$) are those with odd values of $k$, $k=1,3,5...$
	Each oscillator $l$ has its own orthogonal basis. This can be
	easily checked with the help of the Haskell function \emph{bracket},
	the scalar product of two (not normalized) wavefunctions -- each
	annotated by a pair of integers $(l,k)$.

\begin{verbatim}
bracket :: Integer -> (Integer,Integer)
                   -> (Integer, Integer) -> Fraction
bracket n (l, k) (l', k')
    = scalarProduct n f g
    where
        f = wavefunction l k
        g = wavefunction l' k'
\end{verbatim}
        The integer $n$ specifies the space dimension $n=1,2,3..$ when
	computing volume integrals for radial cases. But we can use the same
	function for computing linear integrals as well -- by setting, somewhat
	artificially, this argument to zero, $n=0$. For example

\begin{verbatim}
decimal 8 $ bracket 0 (7,1) (7,1) ==> 14034.40729347
decimal 8 $ bracket 0 (7,1) (7,5) ==> 0
\end{verbatim}
	Notice the convertion from fractional to decimal representation,
	specified here with accuracy to eight decimal places. All internal
	computations are performed on integers of unlimited size and
	on fractions made of such integers. The accuracy can be extremely good,
	depending on the value of \emph{eps}, set at the top of module
	\emph{Spike}. It is only for presentation purposes that we
	convert fractions to decimal representation.

	The following table summarizes the case of linear spiked oscillators

\vspace{2 mm}
\begin{quote}
\begin{quote}
\begin{tabular}{|l|l|l|}
\hline
	$l = 0$         &$k = 0,1,2 ...$           &$E_{0,k} = 1,3,5 ...$   \\
        $l = 1$         &$k = 1,3,5 ...$           &$E_{1,k} = 5,9,13 ...$  \\
        $l = 2$         &$k = 1,3,5 ...$           &$E_{2,k} = 7,11,15 ...$ \\
        ...             &...                    &...\\

\hline
\end{tabular}
\end{quote}
\end{quote}
		  	
% ------------------------------------------------	
\subsection	{Radial normalization}
% ------------------------------------------------
	Scalar product of two basis vectors for radial oscillators can be defined
	by a generic form of improper \emph{volume} integral
\begin{equation}
	\braket {l,i}{l,j} = \int_0^{+\infty} f_{l,i}(x) \, f_{l,j}(x) \, e^{-x^2} \,dV, \label{eq:delta}
\end{equation}
	where $dV$ represents volume of an elementary sphere in N-dimensional space.
	Using general formula for a volume of a sphere with radius $x$ in
	$N$ dimensions
	\[ V(N) = \frac {\pi^{N/2}} {\Gamma (1 + N/2)} x^N, \]
	where
\begin{eqnarray*}
	\Gamma (0)   & = & 1 \\
        \Gamma (1/2) & = & \sqrt {\pi} \\
	\Gamma (1)   & = & 1 \\
	\Gamma (N)   & = & (N - 1) \Gamma (N - 1),
\end{eqnarray*}
	the elementary  volume $dV$, and thus volume integrals (\ref{eq:delta}) can
	be easily computed. Specificly, the `volume' of sphere with radius $x$ in
	dimensions $N=1,2,3$ are given by $2 x, \pi x^2, 4/3\pi x^3$, etc.
	To simplify Haskell computations we will ignore constant factors in
	definition of the N-dimensional volume and define $dV$ simply as
	$dV = dx, xdx, x^2dx$, etc., for $N=1,2,3..$, respectively.

% ------------------------------------------------
\subsubsection {One dimensional}
% ------------------------------------------------

	The `volume' integral in one dimension is twice the value of the one
	dimensional integral over $[0,+\infty)$.
	This is, generally, not the same as the improper integral over
	$(-\infty,+\infty)$ -- unless the function being integrated is even:
	$g(-x)=g(x)$.

	The recurrence formulae for integrals
	$P_n = 2 \int_0^{+\infty} x^n \, e^{-x^2} \, dx$, are given below.
\begin{eqnarray*}
        P_0  & = & \sqrt{\pi} \\
        P_1  & = & 1 \\
	P_n  & = & \frac{n-1}{2} \, P_{n-2}.
\end{eqnarray*}
	As opposed to the case of one dimensional linear normalization
	the integrals $P_n$ for odd powers of $n$ do not vanish.

	The case $l=0$ admits all solutions $k=0,1,2...$, but further investigation
	reveals that they do not form the basis since scalar products of
	one odd and one even eigenfunctions is not equal to zero, as in this example,

	\[ \braket {0,0} {0,1} = 0.797884. \]
	However, the set of solutions for $l=0$ can be split into two sets:
	one even ($k=0,2,4,6...$) and one odd ($k=1,3,5...$) -- each forming
	the basis
	\[ \braket {0,k} {0,k'} = \delta_{kk'} .\]

	Solutions for the remaining oscillators ($l > 0$) are restricted to odd
	eigenfunctions ($k=1,3,5..$), because even eigenfunctions are unphysical.
	As a result we can summarize this case as follows:

\paragraph{Even $k$ basis}
\vspace{2 mm}
\begin{quote}
\begin{quote}
\begin{tabular}{|l|l|l|}
\hline
	$l = 0$         &$k = 0,2,4 ...$           &$E_{0,k} = 1,5,9 ...$ \\
\hline
\end{tabular}
\end{quote}
\end{quote}	

\paragraph{Odd $k$ bases}
\vspace{2 mm}
\begin{quote}
\begin{quote}
\begin{tabular}{|l|l|l|}
\hline
	$l = 0$         &$k = 1,3,5 ...$           &$E_{0,k} = 3,7,11 ...$   \\
        $l = 1$         &$k = 1,3,5 ...$           &$E_{1,k} = 5,9,13 ...$  \\
        $l = 2$         &$k = 1,3,5 ...$           &$E_{2,k} = 7,11,15 ...$ \\
        ...             &...                       &...\\

\hline
\end{tabular}
\end{quote}
\end{quote}

	This table is particularly interesting because it summarizes
	what is traditionally known as \emph{three dimensional isotropic
	harmonic oscillator}\cite{Cohen}. When expressed in spherical coordinates
	and after separation of variables the radial part of its Hamiltonian
	becomes

	\[ H = -\frac{\hbar^2}{2m}\frac{1}{r}\frac{d^2}{dr^2}r +
	\frac{1}{2}m\omega^2r^2 + \frac{l(l+1)\hbar^2}{2mr^2}, \]
	which can be further reduced to the form (\ref{eq:hamiltonians})
	after representing the radial solutions as

	\[R_{l,k} = \frac{1}{r} \phi_{l,k}(r). \]
	
	Functions $\phi_{l,k}$ corespond to $f_{l,k}e^{-r^2/2}$ investigated
	in this paper, and the three dimensional normalization of solutions
	$R_{l,k}$ becomes one dimensional normalization of $\phi_{l,k}$

	\[ \int_0^\infty R_{l,k}R_{l,k'} \,4\pi r^2 \,dr =
           4\pi \int_0^\infty \phi_{l,k}\phi_{l,k'} \, dr. \] 	

	According to (\ref{eq:energies}) energies are given by $E_{l,k}=2(l+k)+1$,
	in units of $\frac{\hbar\omega}{2}$. It is customary to represent this
	as $\varepsilon_n = (n + 3/2)\hbar\omega$, where $n=l+k-1$ -- taking one
	of the values $n=0,1,2,3$...

	With every $n$, we can associate either one or many pairs of $(l,k)$, which
	generate the states $\ket{l,k}$ coresponding to the same energy $\varepsilon_n$

\begin{eqnarray*}
	n = 0  & \Longrightarrow &  \ket{0,1} \\
        n = 1  & \Longrightarrow &  \ket{1,1} \\
        n = 2  & \Longrightarrow &  \ket{0,3}, \, \ket{2,1} \\
	n = 3  & \Longrightarrow &  \ket{1,3}, \, \ket{3,1} \\
	n = 4  & \Longrightarrow &  \ket{0,5}, \, \ket{2,3}, \, \ket{4,1} \\
        ...    & \Longrightarrow &   ...
\end{eqnarray*}

	The energy levels are degenerate; when $n$ is even then there are
	$\frac{n}{2}+1$ states

	\[ \ket{0,n+1}, \, \ket{2,n-1}, \, \ket{4,n-3} ...\]
	and when $n$ is odd then there are $\frac{n+1}{2}$ states

        \[ \ket{1,n}, \, \ket{3,n-2}, \, \ket{5,n-4} ... \]
	associated with each energy level.
	When $2l+1$ possible eigenvalues of $L_3$ are also
	considered for each $l$ then the total degeneracy of energy levels becomes
        $\frac{1}{2}(n+1)(n+2)$. The solutions obtained by other means (
	see for example\cite{Cohen}) are exactly the same as the ones presented in
	this paper.

% ---------------------------------------------------
\subsubsection {Two dimensional}
% ---------------------------------------------------
	According to the condition $N \geq 2l + 1$, we can accept
	both even and odd solutions for oscillator $l=0$, while the even solutions
	for the remaining oscillators ($l > 0$) must be rejected as unphysical.
	However, it seems that the two dimensional normalization does not lead
	to any well defined pattern of orthogonal bases and therefore we
	reject this case as ill-specified. But similar normalization in the next
	dimension ($N=3$) provides us with several usable bases, as shown below.

% --------------------------------------------------
\subsubsection {Three dimensional}
% -------------------------------------------------
	In accordance with the general pattern of physical admissibility of
	even wavefunctions, $N \geq 2l+1$, the first two oscillators,
	$l=0$ and $l=1$, admit both even and odd solutions $k=0,1,2 ...$.
	For the remaining oscillators, $l \geq 2$, only the odd solutions
	$k=1,3,4 ...$ are acceptable.
	
	For each $l$, the following pattern of orthogonality relations emerges

	\[ \braket {l,k} {l,k'} = \delta_{kk'} \qquad \mbox{for } |k'- k|=0,4,8,12... \] 	
	Consequently, the eigenvectors of the first two oscillators can be partitioned
	into four bases - two even and two odd, while each of the remaining oscillators,
	$l \geq 2$, has two orthogonal bases. The following four tables summarize the
	three dimensional normalization.

\paragraph{Even $k$ bases}
\vspace{2 mm}
\begin{quote}
\begin{quote}
\begin{tabular}{|l|l|l|}
\hline
	$l = 0$         &$k = 0,4,8 ...$           &$E_{0,k} = 1,9,17 ...$   \\
        $l = 1$         &$k = 0,4,8 ...$           &$E_{1,k} = 3,11,19 ...$ \\
\hline
\end{tabular}
\end{quote}
\end{quote}

\vspace{2 mm}
\begin{quote}
\begin{quote}
\begin{tabular}{|l|l|l|}
\hline
        $l = 0$         &$k = 2,6,10 ...$          &$E_{0,k} = 5,13,21 ...$  \\
        $l = 1$         &$k = 2,6,10 ...$          &$E_{1,k} = 7,15,23 ...$ \\
\hline
\end{tabular}
\end{quote}
\end{quote}

\paragraph{The first odd $k$ basis}

\vspace{2 mm}
\begin{quote}
\begin{quote}
\begin{tabular}{|l|l|l|}
\hline
	$l = 0$         &$k = 1,5,9 ...$           &$E_{0,k} = 3,11,19 ...$   \\
        $l = 1$         &$k = 1,5,9 ...$           &$E_{1,k} = 5,13,21 ...$  \\
        $l = 2$         &$k = 1,5,9 ...$           &$E_{2,k} = 7,15,23 ...$ \\
 	 ...             &...                       &... \\
\hline

\end{tabular}
\end{quote}
\end{quote}
	Its degenerated energy levels are given by
	
	\[ \varepsilon_n = (n + \frac{3}{2}) \hbar\omega, \]
	where $n=l+k-1=0,1,2...$
\paragraph{The second odd $k$ basis}
	
\vspace{2 mm}
\begin{quote}
\begin{quote}
\begin{tabular}{|l|l|l|}
\hline
	$l = 0$         &$k = 3,7,11 ...$          &$E_{0,k} = 7,15,23 ...$   \\
	$l = 1$         &$k = 3,7,11 ...$          &$E_{1,k} = 9,17,25 ...$   \\
        $l = 2$         &$k = 3,7,11 ...$          &$E_{2,k} = 11,19,27 ...$ \\
 	 ...             &...                       &... \\
\hline

\end{tabular}
\end{quote}
\end{quote}
	Its degenerated energy levels are given by

	\[ \varepsilon_n = (n + \frac{7}{2})\hbar\omega \]
	where $n=l+k-3=0,1,2,3...$
	
% --------------------------------------------------
\subsubsection {N dimensional}
% -------------------------------------------------
	
        Specific cases, examined in the last few sections,
	exhibit certain pattern, which can be generalized on $N$ dimensions
	and directly verified by Haskell function \emph{bracket}.
	
	In each $N$ dimensional space there exists a treshold integer number $l'=(N-1)/2$,
	such that for all
	$l \leq l'$ all wavefunctions $k=0,1,2 ...$ can be accepted as physical.
	Above this threshold the even solutions $k=0,2,4$ must be rejected as
	unphysical.
	However, it appears that radial normalization in any of the even dimensions
	$N=2,4,6...$ does not lead to properly defined basis or bases. In contrary, for each of the odd dimensions
	$N=1,3,5..$ there is a clear pattern of several staggered orthogonal
	bases. Half of them are even in $k$ and they are subjected to the
	limitation of physicality. The other half are orthogonal bases with odd vectors $k$
	and they are all physical and well defined.

	For $N=1$ the vectors belonging to the same basis are enumerated
	by $\Delta k= 2$. For $N=3$ the staggering of bases is given by $\Delta k=4$, for
	$N=5$ it is $\Delta k=6$, etc.

% --------------------------------------------------
\section {Conclusions}
% -------------------------------------------------

	We have shown that spiked oscillators described by Hamiltonians (\ref{eq:hamiltonians})
	have simple and exact eigenfunctions, subject to further normalization
	restrictions, which are specific to a problem dimension. One such restriction is
	related to integrability of the solutions and the other to selection
	of orthogonal bases. The form (\ref{eq:hamiltonians}) can be interpreted
	either as a family of one dimensional linear spiked oscillators,
	or families of radial spiked oscillators in odd $N$ dimensional spaces,
	where $N=1,3,5..$. Even dimensions $N=2,4,6..$ must be rejected because
	they do not give rise to orthogonal bases.

	Each Hamiltonian,$H_l$,
	has an uncountable number of eigenvectors $\ket{l,k}$, where $k=0,1,2..$
	Generally, all solutions indexed by odd $k=1,3,5..$ are integrable and constitute
	one basis or several interleaved bases. In contrary, only a limited
	number of oscillators $l$ admit even eigenfunctions $k=0,2,4..$ for a given
	odd dimension $N=1,3,5..$ -- each forming a basis or a set of interleaved
	bases.
	
 	Each Hamiltonian (\ref{eq:hamiltonians}) leads to an equidistant
	energy spectrum, isomorphic to a spectrum of ordinary quantum oscillator,
	but subjected to restrictions described above. Our factorization
	method which is akin to \emph{SUSY} method, and which relies on two
	kinds of operators: \emph{intertwing operators} $b_l\dag$ and $b_l$,
	and \emph{ladder operators} $a_l\dag$ and $a_l$, is purely algebraic.
	But while the $\emph{SUSY}$ method correlates solutions of
	two Hamiltonians with isomorphic spectra, our method extends this
	approach on infinite set of Hamiltonians. The theory
	is augmented by very simple Haskell program, which
	directly implements these operators, generates the eigenfunctions,
	tests integrability of the solutions and verifies orthogonality conditions
	for wavefunctions enumerated by quantum numbers $l$ and $k$.

\end{document}